\documentclass{emulateapj}

\newcommand{\rsun}{$R_\odot$}
\newcommand{\msun}{$M_\odot$}
\newcommand{\kmps}{km\,s$^{-1}$}
\newcommand{\teff}{$T_{\rm eff}$}

\shorttitle{Extremely Low-Mass White Dwarf in a Close Binary}
\shortauthors{Vennes et al.}

\begin{document}

\title{Discovery of a Bright, Extremely Low-Mass White Dwarf in a Close Double Degenerate System}

\author{S. Vennes$^{1,}$\altaffilmark{2}, J. R. Thorstensen$^{3}$, A. Kawka$^{1}$, P. N\'emeth$^{1,2}$, J. N. Skinner$^{3}$, 
A. Pigulski$^{4}$,\\ M. St\c{e}\'slicki$^{4}$, Z. Ko{\l}aczkowski$^{4}$, P. \'Sr\'odka$^{4}$}
\affil{$^1$ Astronomick\'y \'ustav, Akademie v\v{e}d \v{C}esk\'e republiky, Fri\v{c}ova 298, CZ-251 65 Ond\v{r}ejov, Czech Republic}
\affil{$^3$ Department of Physics and Astronomy, 6127 Wilder Laboratory, Dartmouth College, Hanover, NH 03755-352, USA}
\affil{$^4$ Instytut Astronomiczny, Uniwersytet Wroc{\l}awski, Kopernika 11, 51-622 Wroc{\l}aw, Poland}

\altaffiltext{2}{Visiting Astronomer, Kitt Peak National Observatory, National Optical Astronomy Observatory, 
which is operated by the Association of Universities for Research in Astronomy (AURA) under cooperative 
agreement with the National Science Foundation.}

\begin{abstract}
We report the discovery of a bright ($V\sim13.7$), extremely low-mass white dwarf in a close double
degenerate system. We originally selected GALEX~J171708.5+675712 for spectroscopic follow-up
among a group of white dwarf candidates in an ultraviolet-optical reduced proper-motion diagram.
The new white dwarf has a mass of 0.18\,\msun\ and is the primary component of a close double degenerate system 
($P=0.246137$ d, $K_1 = 288$\,\kmps) comprising a fainter white dwarf secondary with $M_2 \approx 0.9$\,\msun. 
Light curves phased with the orbital ephemeris show evidence of relativistic beaming and weaker ellipsoidal variations. 
The light curves also reveal secondary eclipses (depth $\approx 8$ mmag) while the primary eclipses appear 
partially compensated by the secondary gravitational deflection and are below detection limits.
Photospheric abundance measurements show a nearly solar
composition of Si, Ca, and Fe ($0.1-1\,\odot$), while the normal kinematics suggest a relatively recent formation history. 
Close binary evolutionary 
scenarios suggest that extremely low mass white dwarfs form via a common-envelope phase and possible Roche-lobe overflow.
\end{abstract}


\section{Introduction}

Many new members of the class of extremely low-mass (ELM) white dwarfs were discovered in the 
Sloan Digital Sky Survey (SDSS) and the Luyten surveys (e.g., Luyten Palomar, Luyten Half-Second, New Luyten Two-Tenths),
such as SDSS J1234$-$0228 \citep{lie2004}, LP400-200 \citep{kaw2006}, SDSS J0917+4638 \citep{kil2007}, and NLTT~11748 \citep{kaw2009}. 
These objects are the products of close binary evolution and likely emerged from a
common-envelope phase and possible Roche-lobe overflow \citep{nel2001}. Some ELM white dwarfs are paired with a neutron star companion, e.g., 
the pulsar J1012+5307 \citep{van1996}, implicating a range of initial secondary masses extending to $\ga 8$\,\msun.

We have identified $\sim200$ bright hot subdwarf and white dwarf stars in a joint {\it Galaxy
Evolution Explorer} ({\it GALEX}) ultraviolet (UV) survey and Guide Star Catalogue (GSC)
survey \citep[paper I:][and paper II: Nemeth et al., in preparation]{ven2011}. A reduced proper-motion diagram based on the photographic color
$V\sim V_{\rm GSC}$ ($H_V\equiv V+5\log{\mu}+5$) versus the color index $N_{\rm UV}-V_{\rm GSC}$ 
segregated the hot subdwarf and white dwarf candidates. As part of a second installment of our catalog of UV-selected stars, 
we identified GALEX~J171708.5+675712 (hereafter, GALEX~J1717+6757) as a likely white dwarf with $H_V\approx 12.3$.

We present photometric and spectroscopic observations (Section 2), and a model
atmosphere and periodogram analysis showing GALEX~J1717+6757 to be a new
ELM white dwarf in a close binary and we analyze its photometric properties (Section 3).
In Section 4, we estimate the distance, kinematics, and age of the system and 
discuss some implications for close binary evolutionary scenarios.

\section{Observations}

We observed GALEX~J1717+6757 on UT 2011 January 28 using the Ritchey-Chr\'etien Focus Spectrograph (RC-spec) attached 
to the 4-m telescope at Kitt Peak National Observatory (KPNO).
We employed the T2KA CCD and the KPC-10A grating (316 lines per mm) with a dispersion
of 2.75 \AA\ per pixel in first order and centered on 5300 \AA.  We also inserted the
order-sorting filter WG360 and covered a useful spectral range from 3600 to 7200 \AA.
The slit width was set at $1.5\arcsec$ resulting in a spectral resolution of $\approx5.5$ \AA.
We obtained two consecutive exposures of 420 s each.
A preliminary model atmosphere analysis revealed a relatively hot and subluminous
object in a region of the HR diagram populated with ELM white dwarfs.
We also measured a radial velocity $\sim 300$\,\kmps\ indicating either peculiar kinematics or an orbital motion
characterized by a large velocity amplitude.

We therefore obtained time series spectroscopy from 2011 March 18 to 21, using the 2.4-m Hiltner telescope and 
modular spectrograph (modspec) at MDM Observatory on Kitt Peak.  A $1024^2$ SITe CCD gave
2 \AA\ pixel$^{-1}$ and 3.5 \AA\ resolution from 4660 to 6730 \AA.  We obtained a total of 65 exposures of 
480 to 600 s each, spanning nearly 6~h of hour angle and covering all phases of the orbit, and measured radial velocities
of the H$\alpha$ and H$\beta$ lines using a convolution algorithm \citep{sch1980}. 
We also obtained eight velocities using the same setup 2011 June 21 through 25 to further constrain the ephemeris (Section 3.2).

Next, we obtained Johnson $BVI$ photometric time series on 2011 February 8 and 21, and March 19 and 22, using the
0.6-m telescope at Bia{\l}k\'ow Observatory. The exposure times were 60 s in $V$ and $I$, and 100 s in $B$. 
The data were corrected for second-order extinction effects by fitting a line to the differential magnitude versus 
air-mass diagram and subsequently removing the linear dependence.
Each time series provided approximately 5 hours of continuous
coverage in the three photometric bands.
We also obtained high signal-to-noise ratio white-light time series on 2011 June 2, 3, 4, 6, and 7 with
exposure times of 30~s. The time series provided between 2 to 5 hours of continuous coverage.

\begin{table}
\begin{center}
\caption{Photometry and astrometry.\label{tbl-1}}
\begin{tabular}{lr}
\tableline\tableline
Parameter        & Measurement \\
\tableline\\
GALEX $F_{\rm UV}$ & 13.44$\pm$0.10 mag  \\
GALEX $N_{\rm UV}$ & 13.80$\pm$0.10 mag  \\
SDSS $u$          & 13.548$\pm$0.003 mag \\
SDSS $g$          & 13.432$\pm$0.006 mag \\
SDSS $r$          & 13.441$\pm$0.003 mag \\
SDSS $i$          & 13.701$\pm$0.003 mag \\
SDSS $z$          & 13.987$\pm$0.004 mag \\
2MASS $J$         & 13.61$\pm$0.02 mag \\
2MASS $H$         & 13.69$\pm$0.03 mag \\
2MASS $K$         & 13.64$\pm$0.05 mag \\
                  &             \\
R.A. (J2000) & 17 17 08.865 \\
Dec. (J2000) & +67 57 11.33 \\
$\mu_\alpha\cos{\delta}$ & $-6.7\pm2.2$ mas\,yr$^{-1}$ \\
$\mu_\delta$ & $51.6\pm0.5$ mas\,yr$^{-1}$ \\
\tableline
\end{tabular}
\end{center}
\end{table}

Finally, we collected photometric measurements from the SDSS \citep{aba2009}, 2 Micron All-Sky Survey 
\citep[2MASS,][]{skr2006} and {\it GALEX} survey (Table~\ref{tbl-1}). We corrected the
{\it GALEX} photometry for non-linearity \citep{mor2007} and estimated an error of 0.1 mag in the process.
These measurements are useful in building a spectral energy distribution (SED) and in constraining the 
atmospheric parameters of the white dwarf (Section 3.1).
The SDSS $g$-band photometric measurement shows discrepancy with other photometric measurements and was rejected from the SED.
Table~\ref{tbl-1} lists astrometric measurements from the Third U.S. Naval Observatory CCD Astrograph Catalog \citep{zac2010}
that will be used to determine the kinematical properties of GALEX~J1717+6757 (Section 4). 

\begin{figure*}
\begin{center}
\includegraphics[viewport=0 30 590 560, clip, width=0.65\textwidth]{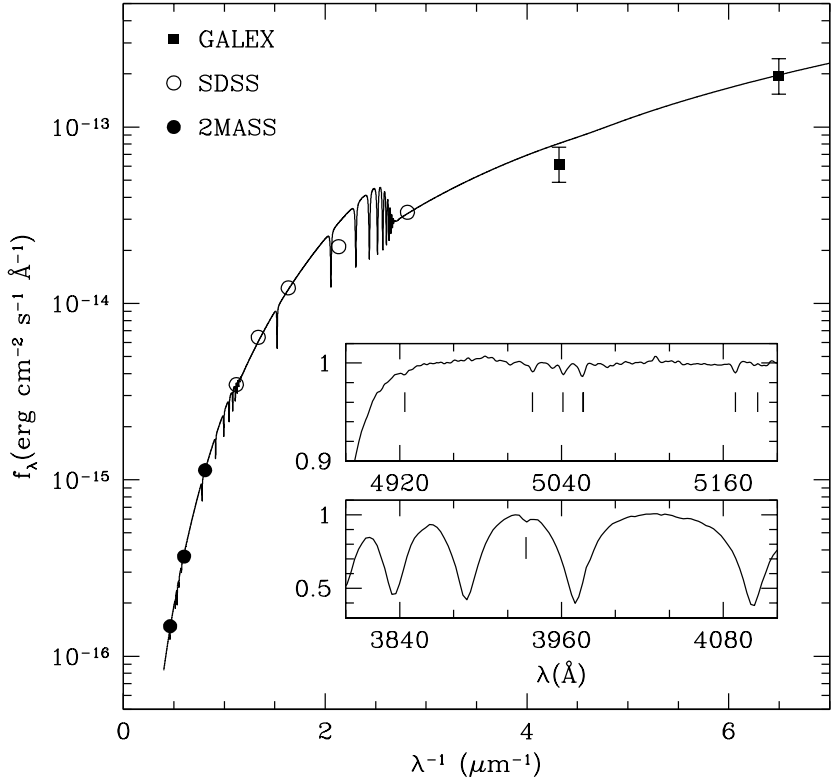}
\end{center}
\caption{Spectral energy distribution (2MASS JHK, SDSS $ugriz$, {\it GALEX} N$_{\rm UV}$ and F$_{\rm UV}$) of GALEX~J1717+6757 compared 
to the best-fitting model 14\,900~K and $\log{g}=5.67$. The effect of minimal extinction ($E_{B-V}=0.005$ mag) is added to the model.
In the inset, the co-added KPNO spectrum shows the upper Balmer lines and a weak photospheric 
\ion{Ca}{2}$\lambda3933$ line, and the co-added MDM spectrum shows \ion{Si}{2}$\lambda\lambda5041,5056,5186$ and
\ion{Fe}{2}$\lambda\lambda4923,5018,5169$ lines. 
\label{fig1}}
\vspace{0.2cm}
\end{figure*}

\section{Analysis}

\subsection{Properties of the primary}

We computed a grid of models in non-local thermodynamic equilibrium (non-LTE) using {\sc Tlusty-Synspec} \citep{hub1995,lan1995} versions 200 and 48,
respectively. 
When fitting the Balmer lines (H$\alpha$ to H13), the best-fitting
parameters
\begin{displaymath}
T_{\rm eff} = 14\,900\pm200 \ {\rm K},\ \ \log{g}=5.67\pm0.05,
\end{displaymath}
and a low helium abundance $\log{[n{\rm (He)}/n{\rm (H)}]} < -2.3$ ($<0.05\,\odot$, where $\odot$ is the solar 
abundance relative to hydrogen) confirmed the nature of GALEX~J1717+6757 as an ELM white dwarf.
Using these parameters we estimated the absolute magnitude in the Sloan $r$ band $M_r = 6.99\pm0.08$,
and with the corresponding distance modulus $r-M_r=6.45\pm0.08$ ($r=13.441$) we
estimated a distance of $195\pm7$ pc.
Figure~\ref{fig1} shows the SED and best-fitting non-LTE model; GALEX~J1717+6757 lies toward a low-density
line-of-sight in the interstellar medium \citep[$E_{B-V}\approx0.03$ mag,][]{sch1998} with negligible extinction in the intervening medium.
The mass and age of the white dwarf may be estimated by locating it on the evolutionary tracks with solar metallicity \citep{ser2001} 
or with low metallicity \citep[$Z=0.001$,][]{ser2002}. The mass is model-dependent and the error is dominated by the choice of
the model grid. Using low-metallicity models we find $M_1=0.184\pm0.001$\msun, but the mass derived from solar-metallicity models 
is less certain because of
a large gap between the 0.169 and 0.196\msun\ sequences. Extrapolating from the sequence of low mass models (0.148, 0.160, 0.169\msun) the
mass is estimated at 0.185\msun. Based on these two model sequences and considering that the ELM lies within
the available sequences, i.e., 0.169 and 0.196\msun, we conservatively estimate $M_1=0.185\pm0.010$\msun\ and $R_1=0.10\pm0.01$\rsun\ for
both low- and high-metallicity models.
We examine the age problem in Section 4.

The surface composition of GALEX~J1717+6757 is that of a heavily polluted white dwarf. Figure~ \ref{fig1} (inset)
shows photospheric heavy element lines that are suitable for abundance measurements. 
Formally classified as DAZ white dwarfs, the hydrogen-rich white dwarfs
with heavy-element spectral lines constitute approximately 20\% of the DA population \citep{zuc2003,koe2005}. 
We fitted the \ion{Ca}{2}$\lambda3933$, \ion{Fe}{2}$\lambda\lambda4923,5018,5169$, and
\ion{Si}{2}$\lambda\lambda$5041,5056,6347,6371 \AA\ lines with high-metallicity non-LTE models computed using {\sc Tlusty-Synspec} and measured:

\begin{eqnarray*}
\log{[n{\rm (Si)}/n{\rm (H)}]} = -5.5\pm0.3\ (0.1\,\odot),\\
\log{[n{\rm (Ca)}/n{\rm (H)}]} = -5.7\pm0.3\ (1.0\,\odot),\\
\log{[n{\rm (Fe)}/n{\rm (H)}]} = -5.1\pm0.3\ (0.3\,\odot).\\
\end{eqnarray*}
The atmospheric composition is somewhat below the solar composition, with a helium
abundance less than 5\% of the solar abundance. High-dispersion and high signal-to-noise ratio spectra 
are required to constrain the abundance of other heavy elements.
The origin of these heavy elements may be fossil
or accreted from the immediate circumstellar environment \citep[see a discussion by][]{koe2006}. 
A direct link to the metallicity of the progenitor cannot be established
because the surface composition of white dwarf stars is determined by diffusion processes 
characterized by time-scales much shorter than evolutionary time-scales.

\subsection{Binary properties and nature of the secondary}

A periodogram constructed from the radial velocities, 
selected an unambiguous period, and a sinusoidal 
fit of the form $v(t) = \gamma + K \sin [2 \pi (t - T_0) / P]$
gave 

\begin{eqnarray*}
T_0 &=& {\rm HJD } \ 2455641.9309 \pm 0.0004,\\
P &=& 0.246137\pm0.000003\ {\rm d},\\
K &=& 288\pm2\ {\rm km\ s^{-1}},\\
\gamma &=& -21.3 \pm 1.7\ {\rm km\ s^{-1}},\\
\end{eqnarray*}
with an RMS residual of only $8$ km s$^{-1}$.
The initial epoch $T_0$ corresponds to the inferior conjunction of the primary.

\begin{figure*}[ht!]
\includegraphics[viewport=0 -10 540 640, clip, width=0.513\textwidth]{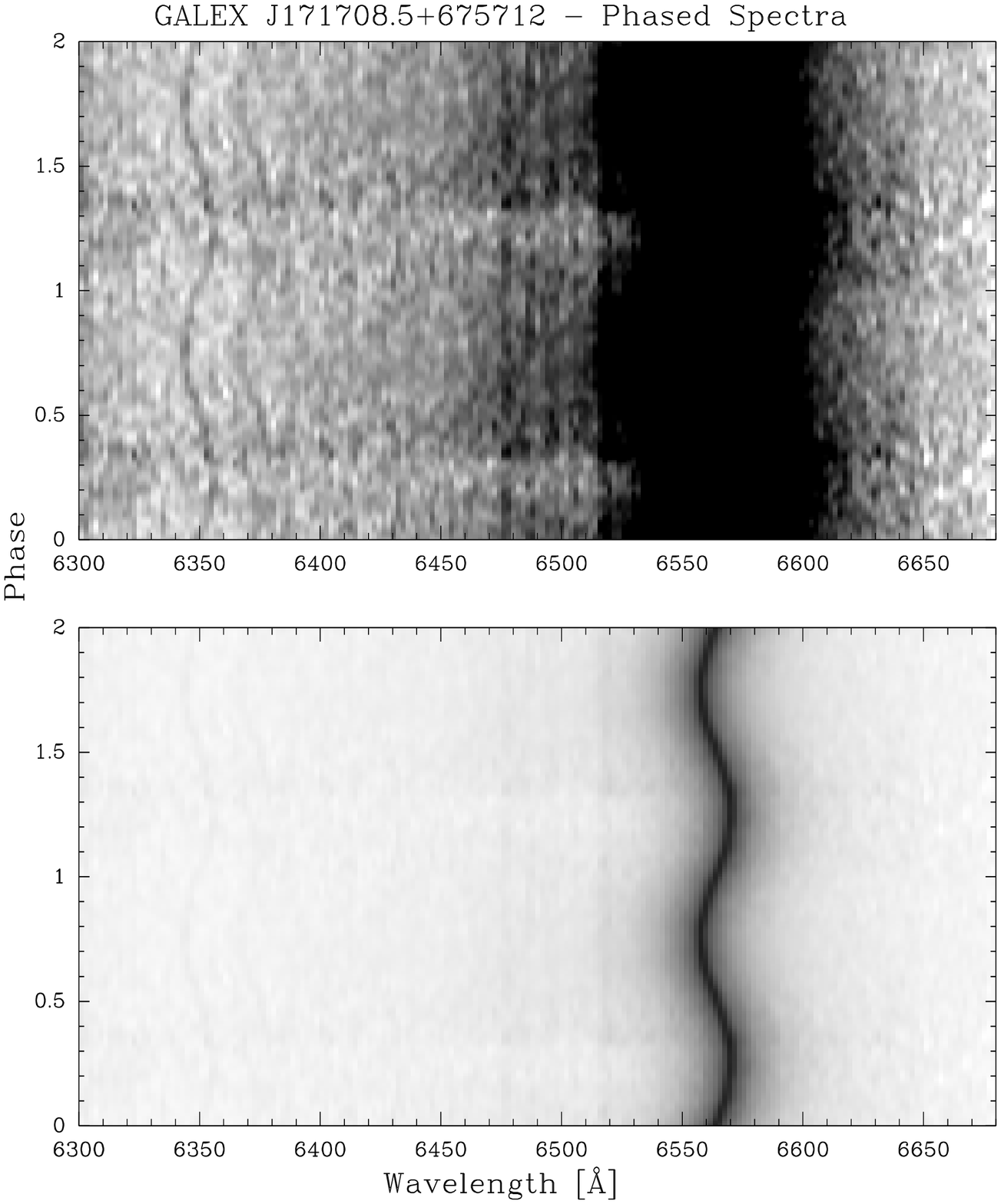}
\includegraphics[viewport=40 0 500 570, clip, width=0.487\textwidth]{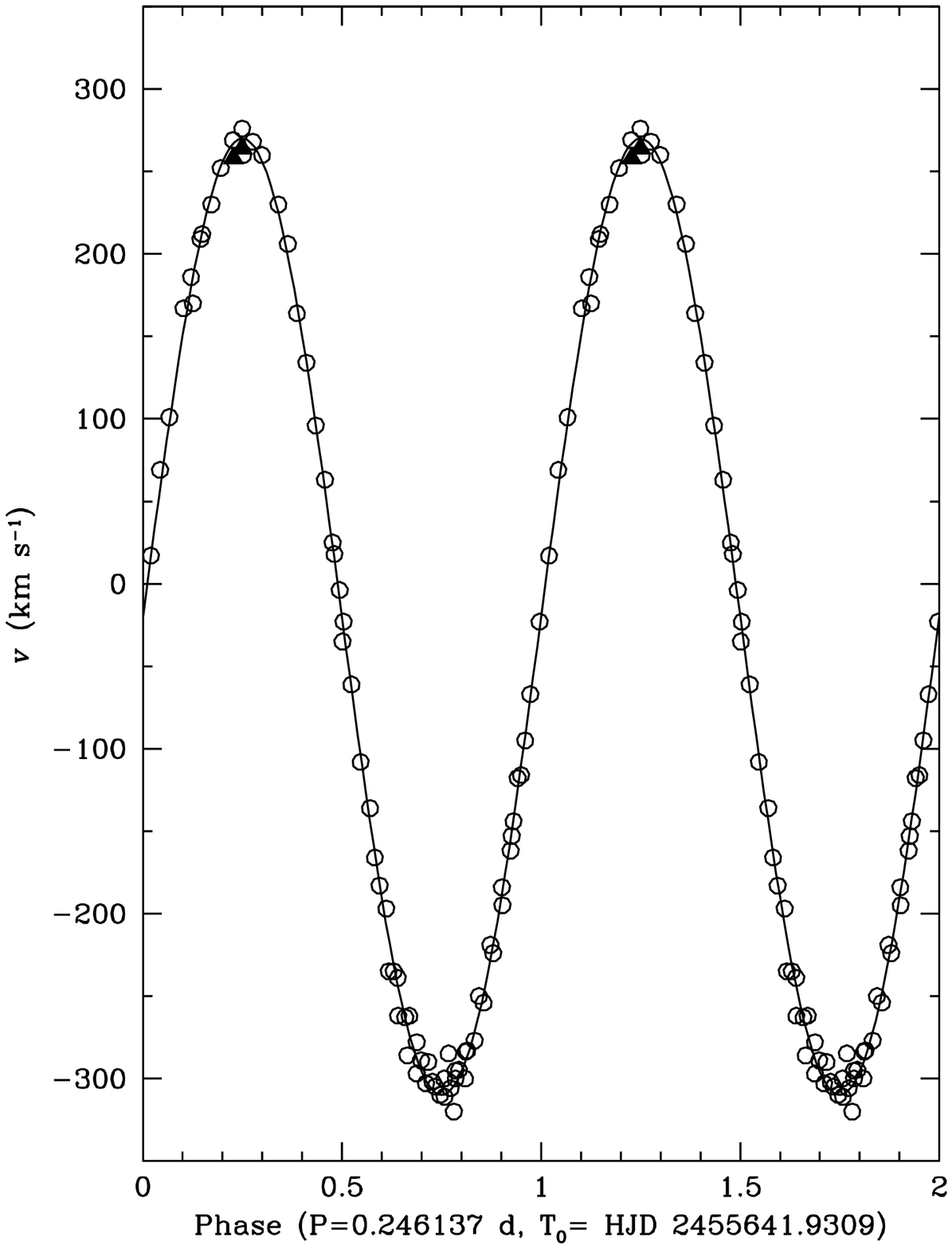}
\caption{(Left) trailed spectra showing \ion{Si}{2}$\lambda\lambda$6347, 6371 and H$\alpha$ lines tracing the orbital
motion (Section 3.2) of the bright ELM white dwarf primary.  The upper panel
is scaled to bring out the \ion{Si}{2} features, and the lower to show the 
core of the H$\alpha$ line. 
(Right) radial velocity measurements phased on the orbital period with the 
best-fitting sinusoidal curve tracing the primary orbital motion.  The data are repeated
for one cycle for continuity.  The full triangles near phase 0.2 are from the Kitt Peak
4 m, and all other velocities are from the MDM 2.4 m.\label{fig2}}
\end{figure*}

Figure~\ref{fig2} shows the spectroscopic trail obtained at MDM Observatory: H$\alpha$ and \ion{Si}{2}$\lambda\lambda$6347, 6371
clearly trace the primary orbit bolstering the photospheric identification of the silicon lines. We found no spectroscopic
signatures of the faint secondary star. The folded radial velocity curve
follows a circular orbit (Fig.~\ref{fig2}).

The period and $K$-velocity imply a mass function for the secondary
\begin{displaymath}
f(M_2) = 0.609\pm0.012\,M_\odot,
\end{displaymath}
or, assuming $M_1=0.185\pm0.010$\,\msun, $M_2\ga 0.86$\,\msun, so the companion must be a compact object.
If the unseen companion is below the Chandrasekhar limit $M_2\la 1.35$\,\msun, then the inclination 
$i\ga60^\circ$.
No radio sources are found in the vicinity of the ELM white dwarf in the NRAO-VLA Sky Survey (NVSS) with
a sensitivity of $\sim0.5$\,mJy at 1400 MHz \citep{con1998}, contrasting with the detection
of the milli-second pulsar PSR~J1012+5307 in the NVSS \citep{kap1998} and other radio observations \citep{nic1995}.
Pulsar companions to ELM white dwarfs are relatively rare \citep{agu2009} and we conclude that the companion
to the ELM white dwarf GALEX~J1717+6757 is a massive white dwarf.

\begin{figure*}
\begin{center}
\includegraphics[viewport=0 0 590 590, clip, width=0.60\textwidth]{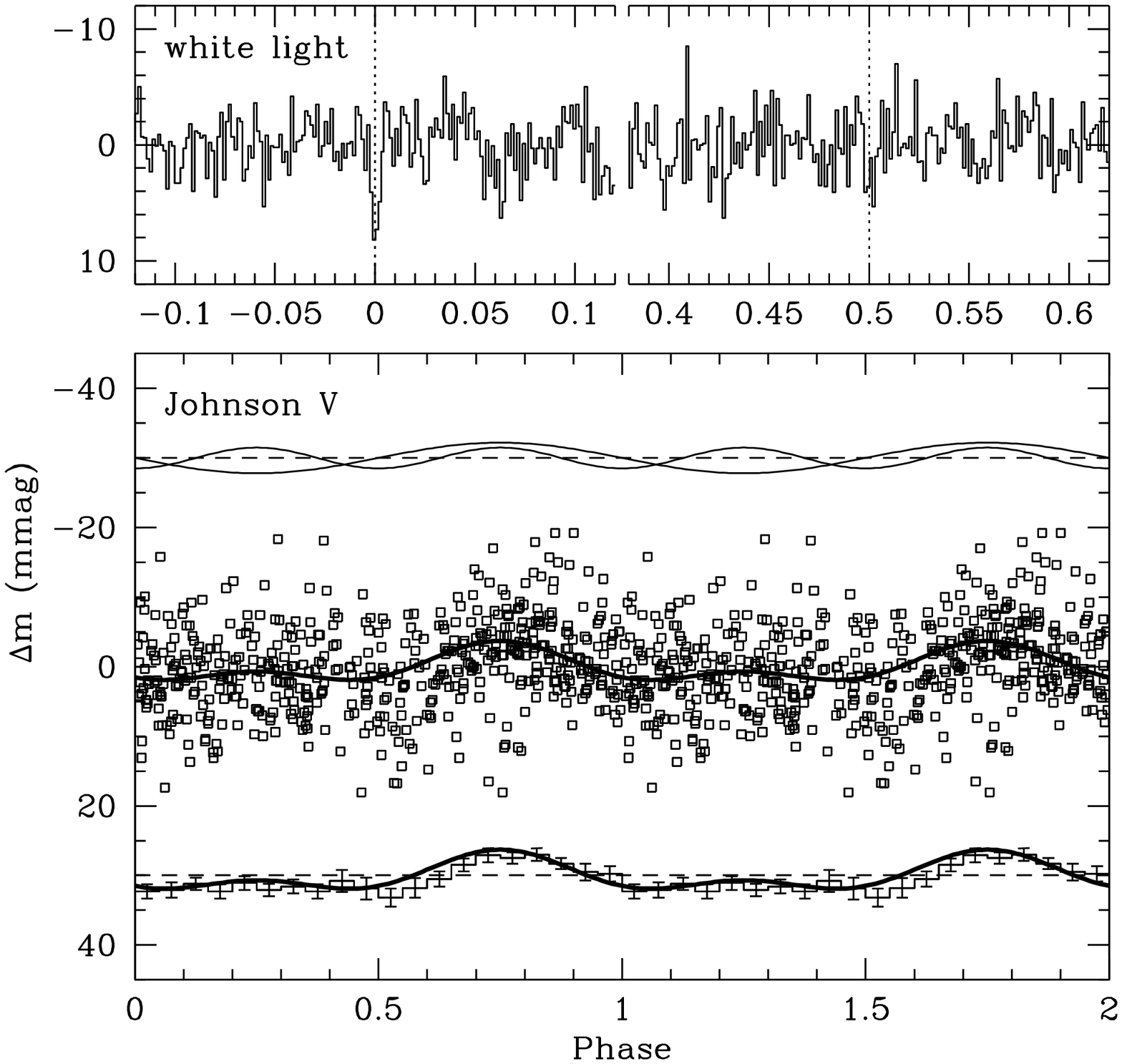}
\end{center}
\caption{(Top panel) white-light time series (30~s bins) folded on the ephemeris (see text) 
covering the primary (phase $=0.5$) and secondary
(phase $=0$) eclipses, and (bottom panel) Johnson $V$ time series (open squares) and model curve (thick line). 
The upper curves in the bottom panel are offset
by $-30$ mmag and depict ellipsoidal effect at half-period and relativistic beaming at full-period (see text),
while the lower curves are offset by 30 mmag and represent the binned (0.05 phase) light-curve (histogram with
error bars) and
the model curve (thick line).
\label{fig3}}
\end{figure*}

\begin{figure*}
\begin{center}
\includegraphics[viewport=0 0 590 570, clip, width=0.600\textwidth]{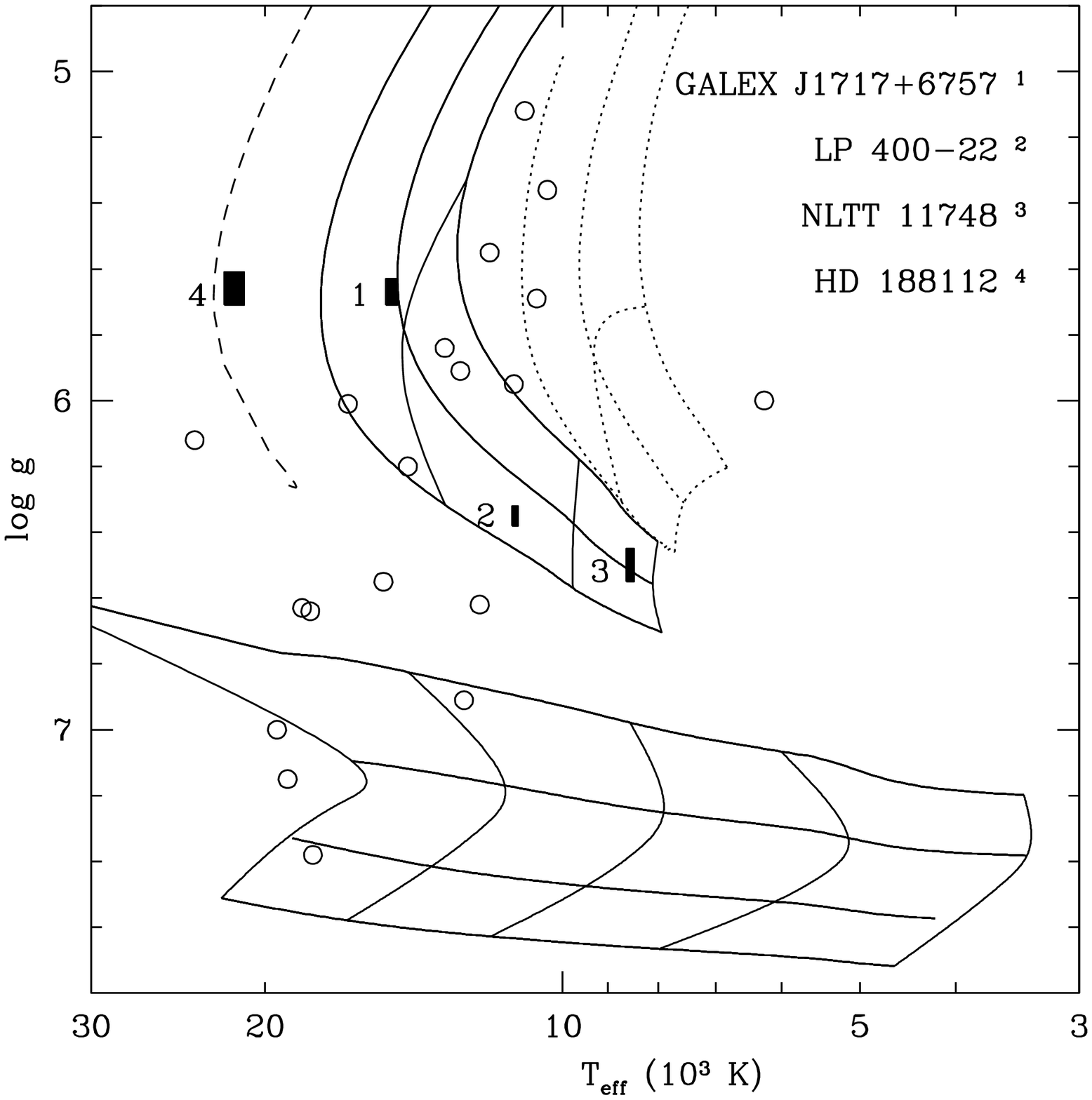}
\end{center}
\caption{Sample of low-mass white dwarfs (open circles, see text) and selected ELM white dwarfs (full rectangles covering
error bars) 
and evolutionary tracks in the (\teff, $\log{g}$) plane. The dashed line shows a 
pre-flash model at 0.234\msun\ \citep{dri1998}. The 
full lines show a series of low-metallicity models at, from top to bottom, 0.172, 0.183 and 0.197\msun\ drawn with isochrones at, from 
right to left, 10, 3, and 1 Gyr, and, from the same grid, models at 0.244, 0.300, 0.380, and 0.449\msun\ with 
isochrones, from right to left, at 10, 3, 1, 0.3, and 0.1 Gyr \citep{ser2002}. The dotted lines show solar-metallicity models at, from top to
bottom, 0.148, 0.160, and 0.169\msun\ with isochrones, from right to left, at 10 and 3 Gyr \citep{ser2001}.\label{fig4}} 
\end{figure*}

We folded the Bia{\l}k\'ow photometric time series on our ephemeris. 
Including all $V$ data, the time series folded on half the period has a weak amplitude of $1.5\pm0.4$ mmag and 
a maximum at phase $0.30\pm0.02$ (statistical error only), i.e. close to phase 0.25 expected for ellipticity 
effect. Using ``JKTEBOP"\footnote[5]{Available at {\tt http://www.astro.keele.ac.uk/}\~{\tt jkt/codes/} {\tt jktebop.html}} \citep{pop1981} 
we verified that the amplitude calculated from the model for the system parameters agrees with the observed amplitude.
The $V$ time series folded on the period has a stronger amplitude of $2.2\pm0.4$ mmag with a maximum at phase 
$0.81\pm0.03$, i.e., also $\approx 2\sigma$ from phase 0.75 expected for 
relativistic beaming. Figure~\ref{fig3} shows the folded $V$ time-series and a model curve combining both effects:
we added stellar distortion with an amplitude of 1.5 mmag and relativistic beaming with an amplitude of 2.2 mmag.
\citet{van2010} and \citet{shp2010} discuss the beaming effect in close binaries; because the primary outshines
the secondary, the amplitude of the effect can be written simply as $A\approx \alpha\,{\rm e}^\alpha/({\rm e}^\alpha-1)\times(K/c)$,
where $c$ is the speed of light, $K$ ($=288$\,\kmps) the velocity semi-amplitude, and $\alpha = h\nu/kT_{\rm eff}$ where $\nu$ is the
frequency of the photometric bandpass and $T_{\rm eff}$ is the stellar effective temperature. In the $V$ band,
the amplitude of the effect in GALEX~J1717+6757 is expected to be $A=2.3$ mmag.
This is the second star in which relativistic beaming 
was detected from the ground after NLTT~11748 \citep{shp2010}.

Adopting for the primary and secondary $M_1=0.185$\msun, $R_1=0.1$\rsun\ (Section 3.1) and $M_2=0.9$\msun, 
$R_2=0.008$\rsun\ \citep[using mass-radius relations of][]{ben1999}, respectively,
the secondary eclipse would last up to 400~s at $i=90^\circ$ with a depth of 7 mmag assuming an identical surface
flux for the primary and secondary stars. 
Employing only the June data (white light) we observed four secondary eclipses at $T({\rm HJD~2455000+}) =
715.5252,\ 716.5098,\ 717.4944$, and $720.4476$, and lasting on average 80-150~s ($i = 86.75\pm0.05^\circ$) with a depth of 8 mmag. 
The eclipse times are consistent with phase $=0$ with an offset of only $-0.003\pm0.004$, 
and suggest only a small adjustment to
the orbital period of $-0.0000025$~d, i.e., within quoted errors. Figure~\ref{fig3} shows the white-light time series
phased with the adjusted period.
The primary eclipse, i.e., the transit of the secondary, is not clearly detected with a maximum depth of $\approx 4$ mmag.
The primary eclipse is partially compensated by gravitational deflection by the massive white dwarf. With the adopted
parameters and a binary separation of 1.7\rsun, the Einstein radius of the secondary is $R_E=0.0036$\rsun. The total
transit depth including the lensing
effect may be written as \citep{mar2001} $\Delta m \approx (R_2/R_1)^2-2(R_E/R_1)^2$, where ``1'' designates the ELM primary, and ``2''
the massive unseen companion. Consequently, the predicted transit is $\Delta m \approx 4$ mmag, consistent with the deepest
feature found at phase 0.5.
This lensing effect has been observed in NLTT~11748 \citep{ste2010}.

\section{Discussion}

The ELM GALEX~J1717+6757 constitutes an interesting test of evolutionary models. Following low-metallicity evolutionary
tracks, the white dwarf has a cooling age of 880$^{+40}_{-30}$ Myr, but it is not possible to derive an age
using the solar-metallicity tracks due to a large gap in properties above 0.169\msun.
Figure~\ref{fig4} compares the evolutionary models to samples of ELM white dwarfs. We selected the old systems
LP~400-22 \citep{kaw2006,kil2009,ven2009}, NLTT~11748 \citep{kaw2009,ste2010,kaw2010} and 
a sample of low-mass white dwarfs in the SDSS survey \citep{kul2010,bro2010,kil2011a,kil2011b}. 
The  ELM precursor HD~188112 \citep{heb2003} is also shown with a pre-flash model describing low-mass subdwarf stars \citep{dri1998}.
The evolutionary tracks cover the population with masses ranging from $\ga 0.15$\msun\ to $\approx0.45$\msun. The ELM
white dwarfs ($M\la0.2$\msun) occupy a relatively high-luminosity region of the diagram characterized by residual
burning. \citet{alt2001} estimates that, taking into account the effect of diffusion, objects with masses $\ga0.18$\msun\ will experience a thermonuclear flash
that delays their cooling. The effect is also metallicity dependent, with higher-metallicity objects experiencing a flash
at a comparatively lower mass. Independent age and mass estimates are possible in binaries comprising an ELM white dwarf
and a pulsar \citep[see][]{alt2001} but are more difficult to obtain in double white dwarf systems. 
The ELM GALEX~J1717+6757 is one of the youngest known class members, but the estimated age of the system remains model-dependent
with the metallicity of the progenitors still unconstrained.

The ELM primary in GALEX~J1717+6757, with an estimated radius $R_1=0.1$\,\rsun\ fills in only about 1.2\% of its Roche volume 
\citep[$R_{1,L} = 0.43$\,\rsun, following][]{egg1983} resulting in low-amplitude ellipsoidal variations compounded with a dominant
relativistic beaming effect. The system is expected to merge ($t_{\rm merge}$) in 7.0~Gyr assuming a
secondary mass $M_2=0.9$\,\msun\ \citep{rit1986}, and
the product of the merger will likely be an ultra-massive white dwarf with $M\approx 1.1$\,\msun.

Using a distance $d=195\pm7$ pc (Section 3.1), and the velocity and proper motion measurements
we computed the ($U,V,W$) velocity vector following \citet{joh1987}:
\begin{displaymath}
(U,V,W) = (-35\pm3, -21\pm3, -8\pm3)\ {\rm km\,s}^{-1}.
\end{displaymath}
The velocity components are characteristic of the thin Galactic disc and similar to the white dwarf population in general \citep{sio1988,pau2006}. We conclude that unlike
the old systems LP~400-22 and NLTT~11748 that show peculiar kinematics, the system GALEX~J1717+6757 is 
kinematically young. The calculation of additional evolutionary tracks covering the observed range of parameters,
particularly between 0.17 and 0.19 \msun,
would benefit studies of ELM white dwarfs and their likely progenitors.

\acknowledgments

S.V. and A.K. are supported by GA AV grant numbers IAA300030908 and IAA301630901, respectively, and by GA \v{C}R grant number P209/10/0967.
A.K. also acknowledges support from the Centre for Theoretical Astrophysics (LC06014). J.R.T. and J.N.S acknowledge
support from NSF grant AST-0708810.  A.P. and Z.K. acknowledge Polish MNiSzW grant N N203 302635.
This research has made use of the VizieR catalogue access tool (CDS, Strasbourg, France), and
of data products from the Two Micron All Sky Survey which is a joint project 
of the University of Massachusetts and the Infrared Processing and Analysis Center/California Institute 
of Technology, funded by the National Aeronautics and Space Administration and the National Science Foundation.

{\it Facilities:} \facility{GALEX}, \facility{Mayall}.

\end{document}